\documentclass{article}
\usepackage{comment}
\usepackage{amsmath,amsfonts}
\usepackage{times}
\usepackage{graphicx} 
\usepackage{subfigure} 

\usepackage{natbib}

\usepackage{algorithm}
\usepackage[noend]{algorithmic}

\usepackage{hyperref}

\usepackage[accepted]{icml2016}

\usepackage{bm}
\usepackage{bbm}
\usepackage{xspace}
\usepackage{ifthen}
\usepackage{tabularx}
\usepackage{mathtools}
\usepackage[inline]{enumitem}

\let\en=\ensuremath

\makeatletter
\def\argopen{\@ifstar{\mathopen{}\mathclose\bgroup}{\mathopen{}\mathclose\bgroup\left}}
\def\argclose{\@ifstar{\egroup}{\aftergroup\egroup\right}}
\makeatother
\newcommand{\myp}[3]{\argopen#1 #3 \argclose#2}

\makeatletter
\newcounter{bal}
\def\scantoken#1#2{\@tfor\next:=#2\do{\ifx\next#1 \stepcounter{bal}\fi}}
\def\p{\@ifstar{\myp@st}{\myp@ns}}
\def\myp@ns{\@ifnextchar({\myp@paren}{%
\@ifnextchar[{\myp@brack}{%
\@ifnextchar\bgroup{\myp@brace}{\relax}
}}}
\def\myp@st{\@ifnextchar({\myp@paren@st}{%
\@ifnextchar[{\myp@brack@st}{%
\@ifnextchar\bgroup{\myp@brace@st}{\relax}
}}}
\def\myp@paren(#1){%
\def\saved{#1}
\scantoken{\lparen}{#1}
\ifnum\thebal>0
  \g@addto@macro\saved{)}
  \expandafter\paren@inner
\else
  \myp{(}{)}{#1}
\fi
}%
\def\paren@inner#1){%
  \addtocounter{bal}{-1}
  \scantoken{\lparen}{#1}
  \g@addto@macro\saved{#1}
  \ifnum\thebal=0 
    \myp{(}{)}{\saved}
  \else 
    \g@addto@macro\saved{)}
    \expandafter\paren@inner
  \fi
}%
\def\myp@brack[#1]{%
\def\saved{#1}
\scantoken{\lbrack}{#1}
\ifnum\thebal>0
  \g@addto@macro\saved{]}
  \expandafter\brack@inner
\else
  \myp{[}{]}{#1}
\fi
}%
\def\brack@inner#1]{%
  \addtocounter{bal}{-1}
  \scantoken{\lbrack}{#1}
  \g@addto@macro\saved{#1}
  \ifnum\thebal=0 
    \myp{[}{]}{\saved}
  \else 
    \g@addto@macro\saved{]}
    \expandafter\brack@inner
  \fi
}%
\def\myp@brace#1{\argopen\lbrace#1\argclose\rbrace}
\def\myp@paren@st(#1){\argopen*(#1\argclose*)}
\def\myp@brack@st[#1]{\argopen*[#1\argclose*]}
\def\myp@brace@st#1{\argopen*\lbrace#1\argclose*\rbrace}
\makeatother

\makeatletter
\newcommand{\DeclareFunction}[2]{%
  \expandafter\newcommand\csname fn@#1@st\endcsname{
    #2\p*}%
  \expandafter\newcommand\csname fn@#1@ns\endcsname{%
    #2\p}
  \expandafter\def\csname#1\endcsname{%
    \@ifstar{\csname fn@#1@st\endcsname}{\csname fn@#1@ns\endcsname}%
  }%
}
\makeatother

\DeclarePairedDelimiter{\abs}{\lvert}{\rvert}
\DeclarePairedDelimiter{\norm}{\lVert}{\rVert}

\let\Exponential=\exp 
\let\Logarithm=\log 
\let\NaturalLog=\ln
\DeclareFunction{exp}{\Exponential}
\DeclareFunction{log}{\Logarithm}
\DeclareFunction{ln}{\NaturalLog}

\newcommand{\setdef}[3][\suchthat]{\left\{#2 #1 #3\right\}}

\newcommand{\superscript}[1]{\en{^{\textrm{#1}}}\xspace}
\renewcommand{\th}[0]{\superscript{th}}
\newcommand{\slfrac}[2]{\left.#1\middle/#2\right.\xspace}
\newcommand{\ILS}[2]{\en{^{#1}\!\!/\!_{#2}}}

\makeatletter
\def\dd{\@ifnextchar[{\@ddorder}{\@ddnormal}}
\def\@ddorder[#1]#2{\mathop{}\!\mathrm{d}^{#1} #2}
\def\@ddnormal#1{\mathop{}\!\mathrm{d} #1}
\makeatother

\renewcommand{\vec}[1]{\en{\bm{\mathrm{#1}}}}

\newcommand{\tvec}[1]{\en{\tilde{\vec{#1}}}}    

\newcommand{\hvec}[1]{\en{\hat{\vec{#1}}}}      

\newcommand{\mat}[1]{\en{{\bm{\mathrm{#1}}}}}

\makeatletter
\def\fronorm{\@ifstar{\fro@star}{\fro@nostar}}
\def\fro@star#1{\norm*{#1}_{\mathsf{F}}}
\def\fro@nostar#1{\norm{#1}_{\mathsf{F}}}
\makeatother

\DeclareMathOperator{\Null}{null}      
\DeclareMathOperator{\trace}{tr}       
\DeclareMathOperator{\Trace}{Tr}
\DeclareMathOperator{\Rank}{rank}      
\DeclareMathOperator{\Diag}{diag}      
\DeclareMathOperator{\Range}{range}
\DeclareMathOperator{\Det}{det}        
\let\Dim=\dim 
\DeclareFunction{dim}{\Dim}            
\DeclareFunction{rank}{\Rank}
\DeclareFunction{tr}{\trace}
\DeclareFunction{Tr}{\Trace}
\DeclareFunction{nul}{\Null}
\DeclareFunction{diag}{\Diag}
\DeclareFunction{range}{\Range}
\DeclareFunction{det}{\Det}

\makeatletter
\def\grad{\@ifnextchar[{\grad@order}{\nabla}}
\def\grad@order[#1]{\nabla^{#1}}
\makeatother

\makeatletter
\newcommand{\pdv}[2]{\begingroup 
  \@tempswafalse\toks@={}\count@=\z@ 
  \@for\next:=#2\do 
    {\expandafter\check@var\next\@nil
     \advance\count@\der@exp 
     \if@tempswa 
       \toks@=\expandafter{\the\toks@\,}%
     \else 
       \@tempswatrue 
     \fi 
     \toks@=\expandafter{\the\expandafter\toks@\expandafter\partial\der@var}}%
  \frac{\partial\ifnum\count@=\@ne\else^{\number\count@}\fi#1}{\the\toks@}%
  \endgroup} 
\def\check@var{\@ifstar{\mult@var}{\one@var}} 
\def\mult@var#1#2\@nil{\def\der@var{#2^{#1}}\def\der@exp{#1}} 
\def\one@var#1\@nil{\def\der@var{#1}\chardef\der@exp\@ne} 
\makeatother

\DeclareMathOperator{\pr}{\mathbb{P}}        
\DeclareMathOperator{\Ev}{\mathbb{E}}       
\def\Pr{\pr\nolimits\p}

\newcommand{\E}[2][]{%
\ifthenelse{\equal{#1}{}}%
{\en{\Ev\argopen[#2\argclose]}}
{\en{\Ev\nolimits_{#1}\argopen[#2\argclose]}}}

\newcommand{\given}{\mathrel{}%
\ifnum\currentgrouptype=16%
\middle\vert%
\else%
\mid%
\fi%
\mathrel{}}%

\DeclareFunction{pp}{p}                 
\DeclareFunction{qq}{q}                 

\newcommand*{\Lap}[1]{\en{\mathsf{Lap}\left(#1\right)}\xspace}         
\newcommand*{\Cat}[1]{\en{\mathsf{Cat}\left(#1\right)}\xspace} 
\newcommand*{\Norm}[1]{\en{\mathcal{N}\left(#1\right)}\xspace}         

\DeclareMathOperator*{\argmin}{arg\,min\,}

\DeclareMathOperator{\Prox}{prox}

\newcommand{\prox}[3][]{
\ifthenelse{\equal{#1}{}}%
{\en{\Prox_{#2}\left(#3\right)}}%
{\en{\Prox_{#1,#2}\left(#3\right)}}%
\xspace}

\newcommand{\I}[2]{\en{\mathbb{I}_{#1}\!\left(#2\right)}\xspace}

\newcommand{\A}[0]{\en{\mathcal{A}}}

\newcommand{\X}[0]{\en{\mathcal{X}}}

\allowdisplaybreaks

\icmltitlerunning{Postprocessing for Iterative Differentially Private Algorithms}

\begin{document} 

\twocolumn[
\icmltitle{Postprocessing for Iterative Differentially Private Algorithms}

\icmlauthor{Jaewoo Lee}{jlee@cse.psu.edu}
\icmladdress{Penn State University,
            University Park, PA 16801}
\icmlauthor{Daniel Kifer}{dkifer@cse.psu.edu}
\icmladdress{Penn State University,
            University Park, PA 16801}

\icmlkeywords{Differential privacy, post-processing,k-means}

\vskip 0.3in
]

\begin{abstract} 
Iterative algorithms for differential privacy run for a
fixed number of iterations, where each iteration learns some
information from data and produces an intermediate output.
However, the algorithm only releases the output of the last iteration,
and from which the accuracy of algorithm is judged. In this paper, we propose a
post-processing algorithm that seeks to improve the accuracy by
incorporating the knowledge on the data contained in intermediate outputs.
\end{abstract} 

\section{Introduction}
\label{sec:intro}
When designing an iterative algorithm for differential privacy, 
fully utilizing the information privately learned from data 
is crucial to the success.
Suppose we have a
simple algorithm $\A$ that calls the function $\mathcal{K}(D, \theta_{t-1})$ $T$
times in a loop and returns $\theta_{T}$ as the final output, where
$D\in \X^N$ is the 
input dataset and $\theta_{t}=\mathcal{K}(D, \theta_{t-1})$ is an
intermediate output at the $t$\th iteration. A large class of machine
learning algorithms, including clustering, classification, and
regression, can be written in this form, where $\mathcal{K}(D,
\theta_t)$ minimizes some objective function and returns $\theta_{t+1}$. 
By the composition theorem~\cite{Dwork2014privacybook}, if the
function $\mathcal{K}$ satisfies 
$\frac{\epsilon}{T}$-differential privacy, the algorithm $\A$ becomes
$\epsilon$-differentially private. 
At each iteration $t$, the algorithm
extracts some information $\theta_t$ from the given dataset $D$ using 
the privacy budget of $\ILS{\epsilon}{T}$, but it only releases the final
output $\theta_T$ (thus, the accuracy of the algorithm is largely
dependent on the magnitude of noise at the final iteration). 

In this paper, we ask the following question: 
``\emph{Can we improve the accuracy of the final output $\theta_T$ by
incorporating the knowledge contained in the intermediate outputs
$\theta_1, \ldots, \theta_{T-1}$?}''
Recent studies have shown that post-processing algorithms that make 
inferences on the original data from noisy outputs can
significantly improve the accuracy of the
results~\cite{Lee2015admm,Hay2010Boosting,Lin2013HMM}. The main source
of improvements comes from enforcing \emph{consistency constraints}, a
set of (hard) syntactic conditions that hold true for the original
data. Inspired by these post-processing algorithms, we view the
intermediate outputs $\theta_1, \ldots, \theta_{T-1}$ as soft
constraints on our estimates (i.e., the original data). Note that the
privacy guarantee of $\A$ is not degraded by this post-processing.
as long as it doesn't rely on the randomness of $\A$.

Consider a differentially private algorithm $\A$ that generates a
sequence of noisy statistics $\{\tilde{\theta}_1, \ldots,
\tilde{\theta}_T\}$ such that $\tilde{\theta}_t = \mathcal{K}(D,
\tilde{\theta}_{t-1}) + Y$, where $Y$ is a random variable
representing the noise added for privacy. Our goal
is to estimate a dataset $\widehat{D}$ from which 
$\tilde{\theta}_1, \ldots, \tilde{\theta}_T$ are most likely to be
generated. Once we have estimated $\widehat{D}$, a new estimator
$\hat{\theta}$ can be obtained by repeatedly running $\mathcal{K}$ on 
$\widehat{D}$ without noise (contrast this to $\tilde{\theta}_T$
produced with noise and using fixed number of iterations).
Informally, we try to find 
$\widehat{D}$ such that  $\mathcal{K}(D, \tilde{\theta}_t) \approx
\mathcal{K}(\widehat{D}, \tilde{\theta}_t)$ for $t=1, \ldots,
T$. We note that the size of $\widehat{D}$ could be different from
that of the original dataset, $N$; we only require intermediate outputs
of $\mathcal{K}$ on both datasets are similar. However, it is still
challenging to efficiently explore the space of all 
possible datasets. To this end, we propose to use MCMC method with
carefully designed proposal distribution. 
The proposed algorithm builds a Markov chain over the space of all
possible datasets and makes use of noisy statistics to efficiently
propose the next state.
Given a dataset $D_t$, the proposed algorithm samples a new dataset 
$D'$ and determines whether to accept or reject the dataset by
considering the ratio of $\mathbb{P}(\tilde{\theta}_1 \ldots,
\tilde{\theta}_T \given D')$ to $\mathbb{P}(\tilde{\theta}_1, \ldots,
\tilde{\theta}_T \given D_t)$, i.e., Metroplis-Hastings step. 
While doing so, it keeps track of the best scoring dataset.

In this paper, we instantiate this post-processing algorithm in the
context of $K$-means clustering.
The contributions of this paper are as follows:
\begin{itemize}[topsep=0pt,noitemsep,leftmargin=*]
\item We propose a general framework for post-processing a sequence of
  noisy private outputs, which improves the accuracy by incorporating
  intermediate results into the process.
\item We applied our framework to $K$-means clustering problem and
  introduce an efficient proposal distribution that yields low rejection rate.
\item Extensive empirical evaluations on both synthetic and real
  datasets are provided to validate our proposed approach. 
\end{itemize}



\section{Related Works}
\label{sec:related}
We discuss differentially private algorithms that can
be applied to the $K$-means problem. 
The first algorithm is DP-KMEANS introduced
in~\cite{Blum2005SULQ,McSherry2009PIQ}. Each step of the algorithm is
descripbed in Algorithm~\ref{alg:dpkm}.
\begin{algorithm}[tb]
  \caption{DP-KMEANS algorithm}
  \label{alg:dpkm}
\begin{algorithmic}
   \STATE {\bfseries Input:} data $D$, \# of clusters $K$,
   \# of iterations $T$ 
   \STATE Initialize $\vec{c}_1^{(0)}, \cdots, \vec{c}_K^{(0)}$
   \FOR{$t=1$ {\bfseries to} $T$}
     \FOR{$j=1$ {\bfseries to} $K$}
       \STATE
       $B_j^{(t)}=\setdef[:]{\vec{x}_i}{j=\argmin_k\,\norm{\vec{x}_i-\vec{c}_k}_2^2}$ 
       \STATE $\tilde{n}_j^{(t)} \gets |B_j^{(t)}| + \Lap{\frac{2T}{\epsilon}}$
       \STATE $\tvec{s}_j^{(t)} \gets \left(\sum_{\vec{x}_i \in B_j}
         \vec{x}_i\right) + \Lap{\frac{2T}{\epsilon}}^d$ 
       \STATE $\vec{c}_j^{(t)} \gets \slfrac{\tvec{s}_j^{(t)}}{\tilde{n}_j^{(t)}}$
     \ENDFOR
   \ENDFOR
\end{algorithmic}
\end{algorithm}
%
%
The algorithm is almost identical to its non-private counterpart,
Lloyd's algorithm, with two differences. 
First, the algorithm takes a positive integer $T$ as input and is only
run for $T$ iterations. This is to split the given privacy budget
$\epsilon$ into each iteration.
Second, the centroid update 
is done by using noisy sum and noisy count. The use of noisy
statistics generated by the Laplace mechanism ensures that each update
is differentially private. 
 
It is easy to see that the sensitivity of $(n_1^{(t)}, \ldots,
n_K^{(t)})$ is 1 as adding or removing one data point can change the
size of one cluster by 1. Assuming $\X$ is the unit $L_1$-ball (i.e.,
$\norm{\vec{x}_i}_1\leq 1$), the sensitivity of $(\vec{s}_1^{(t)},
\ldots, \vec{s}_K^{(t)})$ is also 1. Therefore, 
together with the argument of composition theorem, 
adding $\Lap{\frac{2T}{\epsilon}}$ to sum and
count 
ensures each iteration satisfies $\ILS{\epsilon}{T}$-differential
privacy. 


GUPT~\cite{Mohan2012GUPT} is a general-purpose system that implements 
the ``sample and aggregate'' framework~\cite{Nissim2007Smooth}. 
Let $f$ be a function on a database. 
In the context of this work, $f$
is the $K$-means clustering algorithm, which takes a database as input
and returns $K$ centroids.
Given a dataset $D$, GUPT first partitions $D$ into
$\ell$ disjoint blocks, say $T_1, \ldots, T_\ell$, and applies $f$ on
each  block $T_i$. The final output of GUPT is computed by averaging
the outputs $f(T_i)$ from each block and adding the Laplace noise to
the average to ensure privacy. 

PrivGene~\cite{Zhang2013PrivGene} is a genetic algorithm based
framework for differentially private model fitting. 
Starting from a
set of randomly chosen solutions, it iteratively improves the quality
of candidate solutions. To be specific, 
the algorithm starts with a
candidate parameter set $\Omega$, initialized with random vectors. At
each iteration, $\Omega$ is enriched by adding offsprings (new
candidate parameters), generated using crossover and mutate operations
on existing parameters. Then, the algorithm selects and maintains a
fixed number of parameters with best fitting scores using exponential
mechanism. 

Recently, Su et al. proposed EUGkM~\cite{Su2015KM}, a non-interactive
grid based algorithm for $K$-means clustering. The main idea is to
divide multi-dimensional space into $M$ rectangular
grid cells. For each grid cell, it releases a pair $(c_i, n_i)$ using
the Laplace mechanism, where $c_i$ and $n_i$ are the center and the
noisy count of data points in the cell, respectively. Note that noise
is only added to the count $n_i$ as releasing $c_i$ has no privacy
implication.
Given a set of pairs $\mathcal{S}=\{(c_i,n_i) : i=1,\ldots, M\}$,
EUGkM considers there are $n_i$ data points 
at $c_i$, and it applies (non-private) $K$-means algorithm on
$\mathcal{S}$. 
They also proposed hybrid method which combines EUGkM with DP-KMEANS.


\section{Postprocessing for K-means}
\label{sec:algorithm}
In this section, we describe the proposed post-processing framework in the context
of $K$-means where the algorithm releases a sequence of noisy cluster
sums and sizes.

\subsection{Inference on Centroids}
Given the $K$ initial centroids $\vec{c}_1^{(0)}, \ldots,
\vec{c}_K^{(0)}$ (chosen independent of $D$),
let $\mathcal{S} = \langle\tilde{\theta}_1, \ldots, \tilde{\theta}_T\rangle$ be the
sequence of (noisy) outputs generated by running Algorithm~\ref{alg:dpkm} for $T$
iterations, where $\tilde{\theta}_t=(\tvec{s}_1^{(t)}, \ldots, \tvec{s}_{K}^{(t)},
  \tilde{n}_1^{(t)}, \ldots, \tilde{n}_K^{(t)})$ for $t=1,
\cdots, T$. Notice that (noisy) cluster centroids $\vec{c}_1^{(t)}, \ldots,
\vec{c}_K^{(t)}$ are completely determined by $\tilde{\theta}_t$. We
abuse notation and use $\tilde{\theta}_t$ to denote both noisy
statistics and $K$ centroids at iteration $t$.
Let $S(D,\theta)$ and $N(D,\theta)$ be the functions
that return the sum and the number of data points in each partition
determined by the given centroids $\theta$. 

Our goal is to make an inference on the cluster centroids
$\vec{c}_1, \ldots, \vec{c}_K$ based on the information $\mathcal{S}$
we learned privately from $D$. We do this by simulating datasets and
evaluating the likelihood of the observed noisy statistics
$\mathcal{S}$ under each dataset. 
Once a dataset that maximizes the likelihood of $\mathcal{S}$ is
found, new estimates for the cluster centroids can be derived by
running a non-private $K$-means algorithm (possibly with multiple random
restarts) on the dataset.
The log-likelihood 
is defined by:
\begin{align*}
&\ln \mathbb{P}[\mathcal{S} \given D] 
=\ln \mathbb{P}[\tvec{s}_1^{(1)}, \ldots, \tvec{s}_K^{(T)}, \tilde{n}_1^{(1)}, \ldots,
  \tilde{n}_K^{(T)} \given D] \\
&=\sum\nolimits_{t=1}^T\left(\ln \Pr[\tvec{s}_1^{(t)},\ldots,\tvec{s}_K^{(t)} \given
  S(D,\tilde{\theta}_{t-1})]\right. \\
&\qquad\qquad + \left. \ln \Pr[\tilde{n}_1^{(t)}, \ldots,
  \tilde{n}_K^{(t)} \given   N(D,\tilde{\theta}_{t-1})]
  \right) \\
&\propto \sum_{t=1}^T\sum_{k=1}^K \norm*{\tvec{s}_k^{(t)} -
  S_k(D,\tilde{\theta}_{t-1})}_1 + \norm*{\tilde{n}_k^{(t)} -
  N_k(D,\tilde{\theta}_{t-1})}_1 \,,
\end{align*}
where the subscript $k$ in
$S_k(D,\tilde{\theta}_{t-1})$ and $N_k(D, \tilde{\theta}_{t-1})$
represent the sum and number of data points in the $k$\th cluster,
respectively.

\subsection{Imposing Consistency}
The accuracy of noisy output $\mathcal{S}$ can be improved by
imposing consistency constraints, using the algorithm proposed
in~\cite{Lee2015admm}. Suppose $\hvec{s}_k^{(t)}$ and
$\hat{n}_k^{(t)}$ are new estimates for $\tvec{s}_k^{(t)}$ and
$\tilde{n}_k^{(t)}$, respectively. It is clear that they should
satisfy the following constraints:
\begin{align*}
& \sum\nolimits_{k=1}^K \hvec{s}_k^{(1)} = \sum\nolimits_{k=1}^K \hvec{s}_k^{(2)} =
\ldots = \sum\nolimits_{k=1}^K \hvec{s}_k^{(T)}\,,\\
& \sum\nolimits_{k=1}^K \hat{n}_k^{(1)} = \sum\nolimits_{k=1}^K \hat{n}_k^{(2)} =
\ldots = \sum\nolimits_{k=1}^K \hat{n}_k^{(T)}\,,\text{ and} \\
& \hat{n}_k^{(t)} \geq 0 \text{ for all } k=1, \ldots, K \text{ and } t=1,
\ldots, T\,.
\end{align*}
For clear semantics and better readability, in the following we
continue to use the notation $\tvec{s}_k^{(t)}$ 
and $\tilde{n}_k^{(t)}$, but they represent the post-processed values.

\subsection{Simulation via MCMC}
The proposed algorithm makes use of approximate sampling method to
find a dataset under which the likelihood of $\mathcal{S}$ is 
maximized. Using MCMC, it samples datasets from the approximate
posterior distribution $\Pr(D \given \mathcal{S})$ and evaluates the
likelihood, while keeping track of the best solution. The target
distribution is 
\[
\begin{aligned}
\pi(D) &= \Exponential\left(-\sum\nolimits_{t=1}^T\sum\nolimits_{k=1}^K
\epsilon_{\!_{S}}\norm{\tvec{s}_k^{(t)} - 
  S_k(D,\tilde{\theta}_{t-1})}_1\right.\\
&\qquad \qquad + \left.\epsilon_{\!_{N}}\norm{\tilde{n}_k^{(t)} -
  N_k(D,\tilde{\theta}_{t-1})}_1\right)\,,
\end{aligned}
\]
where $\epsilon_{\!_{S}}$ and $\epsilon_{\!_{N}}$ correspond to the
privacy budgets for noisy cluster sums and sizes.\footnote{For
  simplicity, we assume $\epsilon_{\!_{S}} = \epsilon_{\!_{N}}$.}
\paragraph*{Proposal distribution}
The Metropolis-Hastings (MH) algorithm requires choice of proposal
distribution, and the convergence of Markov chain to its stationary
distribution $\pi$ is greatly dependent on that choice. The use of a
proposal distribution that is far from $\pi$
will have a high rejection rate and result in slow convergence.

It is shown that $K$-means algorithm can be thought as a special case of
Gaussian Mixture Model (GMM), with means equal to centroids and a common
covariance set to $\delta\mat{I}$ for small $\delta>0$. Given
$\theta_t = (\tvec{s}_1^{(t)}, \ldots,
\tvec{s}_K^{(t)},\tilde{n}_1^{(t)}, \ldots, \tilde{n}_K^{(t)})$, our
proposal distribution is defined to be a mixture of Gaussians:
\begin{equation}
q(\vec{x}) = \sum_{k=1}^K \omega_k\Norm{\vec{x}; \vec{c}_{k}^{(t)},
  \delta\mat{I}}\,,
\label{eq:proposal}
\end{equation}
where $\omega_k = \slfrac{\tilde{n}_k^{(t)}}{\sum_{i=1}^K
  \tilde{n}_i^{(t)}}$ and $\vec{c}_k^{(t)} = \tvec{s}_{k}^{(t)} /
\tilde{n}_k^{(t)}$.  

Given the current dataset $D^{(\tau)}$, a new dataset $D'$ is proposed
by randomly choosing a data point $\vec{x}_i$ from $D$ and replacing
it with a new point $\vec{x}'$ sampled from the proposal
distribution $q$. The sampling of $\vec{x}'$ is done as follows:
\begin{enumerate}[label=(\roman*),leftmargin=*,topsep=1pt,itemsep=-.5ex]
\item choose an integer $t$ randomly from $\{1, 2, \ldots, T\}$.
\item sample $z\given t \sim \Cat{K, \omega_1, \ldots,
    \omega_K}$.
\item sample $\vec{x}'\given z, t \sim \prod_{k=1}^K
  \Norm{\vec{c}_k^{(t)}, \delta\mat{I}}^{\I{}{z=k}}$. 
\end{enumerate}
In the above, $\Cat{K, \omega_1, \ldots, \omega_K}$ represents the
Categorical distribution having possible values in $\{1, \ldots, K\}$,
each with probability mass $\omega_k$ for $k=1,\ldots, K$. $\I{}{z=k}$
is an indicator function whose value is 1 if $z=k$ and 0 otherwise.
\paragraph*{MH Algorithm}
Given $D^{(\tau)}$, the proposed dataset $D'$ of next state
$(\tau+1)$ is accepted with probability
\[
A(D^{(\tau)}, D') = \min\left\{\frac{\pi(\mathcal{S}\given
    D')}{\pi(\mathcal{S}\given D^{(\tau)})}
    \frac{q(\vec{x} \given \vec{x}')}{q(\vec{x}' \given \vec{x})},
    1\right\}\,.
\]
Without loss of generality, suppose a data point $\vec{x}$ is removed
from the $i$\th cluster and a new data point $\vec{x}'$ is added to
the $j$\th cluster at time $\tau$. Then we have
\begin{align*}
&\ln\pi(\mathcal{S}\given D') -\ln\pi(\mathcal{S} \given D^{(\tau)}) \\
&= -\sum_{t=1}^T
\norm{\tvec{s}_i^{(t)} - S_i(D',\tilde{\theta}_{t-1})}_1 -
\norm{\tvec{s}_i^{(t)} - S_i(D^{(\tau)},\tilde{\theta}_{t-1})}_1 \\
& \quad +
\norm{\tvec{s}_j^{(t)} - S_j(D',\tilde{\theta}_{t-1})}_1 - 
\norm{\tvec{s}_j^{(t)} - S_j(D^{(\tau)},\tilde{\theta}_{t-1})}_1 \\
& \quad + 
\abs{\tilde{n}_i^{(t)} - N_i(D',\tilde{\theta}_{t-1})} -
\abs{\tilde{n}_i^{(t)} - N_i(D^{(\tau)},\tilde{\theta}_{t-1})} \\
& \quad + 
\abs{\tilde{n}_j^{(t)} - N_j(D',\tilde{\theta}_{t-1})} -
\abs{\tilde{n}_j^{(t)} - N_j(D^{(\tau)},\tilde{\theta}_{t-1})}\,. 
\end{align*}
We note that $S(D',\tilde{\theta}_{t-1})$ and $N(D',
\tilde{\theta}_{t-1})$ can be calculated from $S(D^{(\tau)},
\tilde{\theta}_{t-1})$ and $N(D^{(\tau)}, \tilde{\theta}_{t-1})$,
respectively. 
The MH correction term is given by
\begin{align*}
\frac{q(\vec{x})}{q(\vec{x}')} 
&= \frac{\qq(\vec{x} \given \vec{z},
  t)\qq(\vec{z}=i)\qq(t)}{\qq(\vec{x}'\given \vec{z}',
  t')\qq(\vec{z}'=j)\qq(t')}  
= \frac{\tilde{n}_i^{(t)}\mathcal{N}(\vec{x}; \vec{c}_i^{(t)},
    \delta\mat{I})}{\tilde{n}_j^{(t')}\mathcal{N}(\vec{x}';\vec{c}_j^{(t')}, \delta\mat{I})}\,.
\end{align*}
The initial state $D^{(0)}$ is initialized with $K$ centroids at the
last iteration. It consists of $\tilde{n}_k^{(T)}$ data points at
$\vec{c}_{k}^{(T)}$ for $k=1, \ldots, K$.



\section{Experiments}
\label{sec:experiment}
In this section, the performance of the proposed post-processing
algorithm is evaluated over both synthetic and real datasets. We note
that our goal is not to develop a better private algorithm for
$K$-means; rather, we seek to improve the accuracy of iterative
differentially private algorithms in general by taking intermediate
results into account.  
Given $K$ partitions $\beta_1, \ldots, \beta_K$ and their centroids
$\vec{c}_1, \ldots, \vec{c}_K$, the quality of 
clustering is measured by the sum of squared distance
between data points and their nearest centroids, within cluster sum of
squares (WCSS); it is the objective function of $K$-means problem.
\[
WCSS = \sum_{k=1}^K \sum_{\vec{x} \in \beta_k} \norm{\vec{x} - \vec{c}_k}_2^2
\]

For the experiments, we used 6 external datasets. 
For all datasets, the domain of each attribute is normalized to $[-1,
1]$ and then projected onto $L_1$-ball.
The characteristics of datasets used in our experiments are
summarized in Table~\ref{tbl:datasets}.
\begin{table}[t]
\caption{Datasets}
\label{tbl:datasets}
\vskip 0.15in
\begin{center}
\begin{small}
\begin{sc}
  \begin{tabular}{cccc}
    \hline
    \abovespace\belowspace
    Datasets & size $N$ & dimension $d$ & $K$ \\
    \hline
    \abovespace
    S1 & 5,000 & 2 & 15 \\ 
    TIGER & 16,281 & 2 & 2 \\
    Gowalla & 107,021 & 2 & 5 \\
    Image & 34,112 & 3 & 3 \\
    Adult (numeric) & 48,842 & 6 & 5\\
    Lifesci & 27,733 & 10 & 3\\
    \hline
  \end{tabular}
\end{sc}
\end{small}
\end{center}
\end{table}
%
For each dataset, we run DP-KMEANS (DPKM) and the proposed method
(MCMC) 10 times and report the averaged WCSS. 
The performance of
$K$-means algorithm is largely dependent on the choice of initial
centroids, it is important to carefully select them. As
in~\cite{Su2015KM}, $K$ initial centroids are chosen independent of
data such that pairwise distance between centroids are greater than
some given constant.

Throughout the experiments, the number of iterations $T$ for DPKM is fixed
to 5. For the proposed algorithm, the length of Markov chain is fixed
to 30,000 and the value of $\delta$, the variance of Gaussian
component in the proposal distribution, is set to 0.001.
\begin{figure*}
  \begin{center}
    \subfigure[S1]{%
      \includegraphics[width=.3\textwidth]{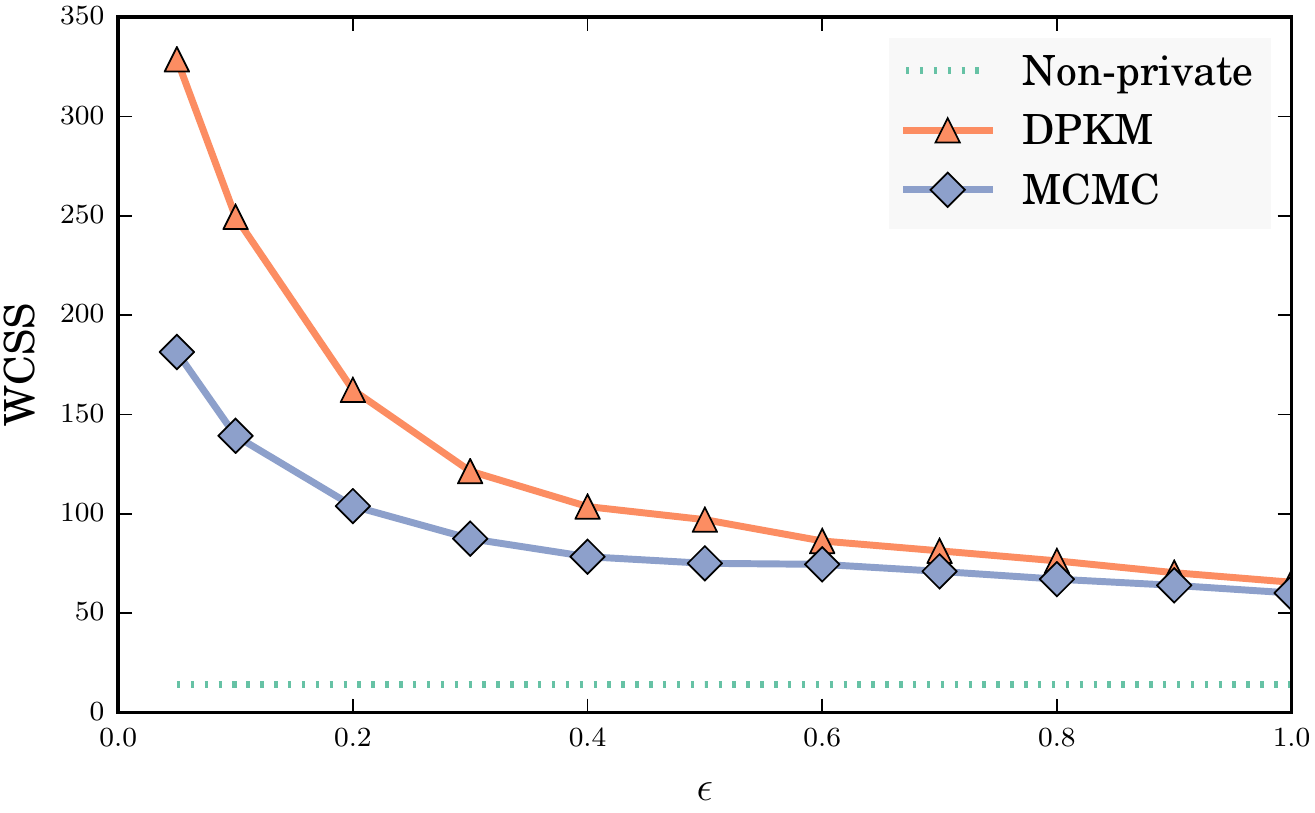}
    }
    \subfigure[Adult-num]{%
      \includegraphics[width=.3\textwidth]{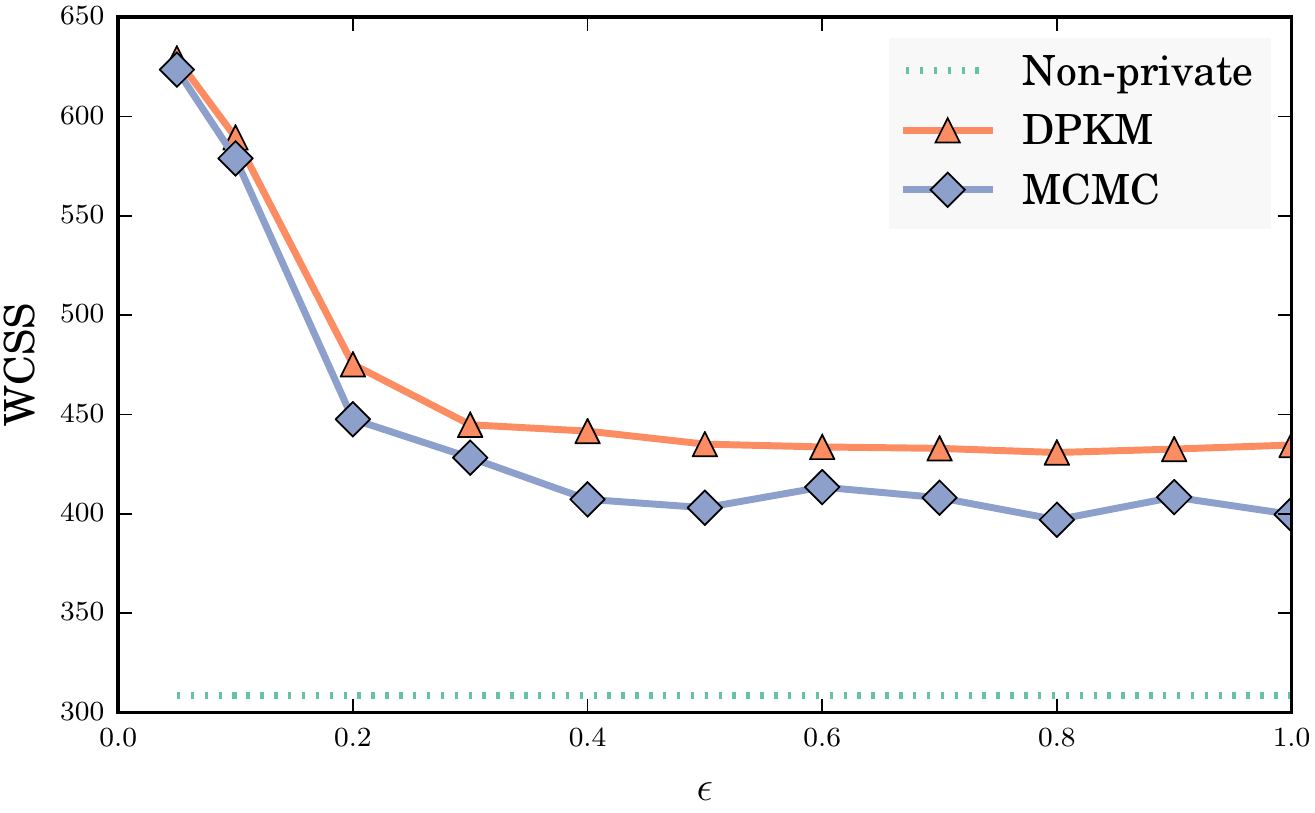}
    }
    \subfigure[Tiger]{%
      \includegraphics[width=.3\textwidth]{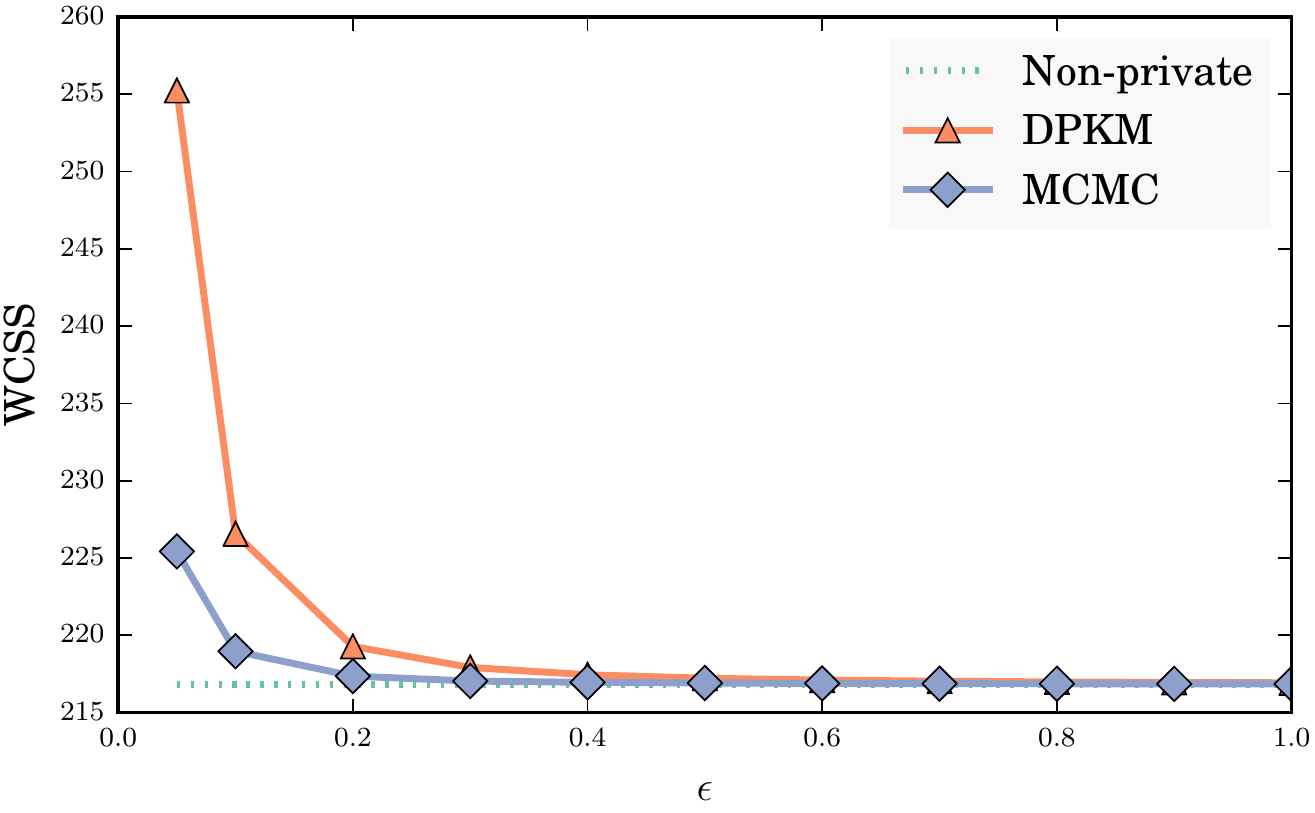}
    }
    \subfigure[Image]{%
      \includegraphics[width=.3\textwidth]{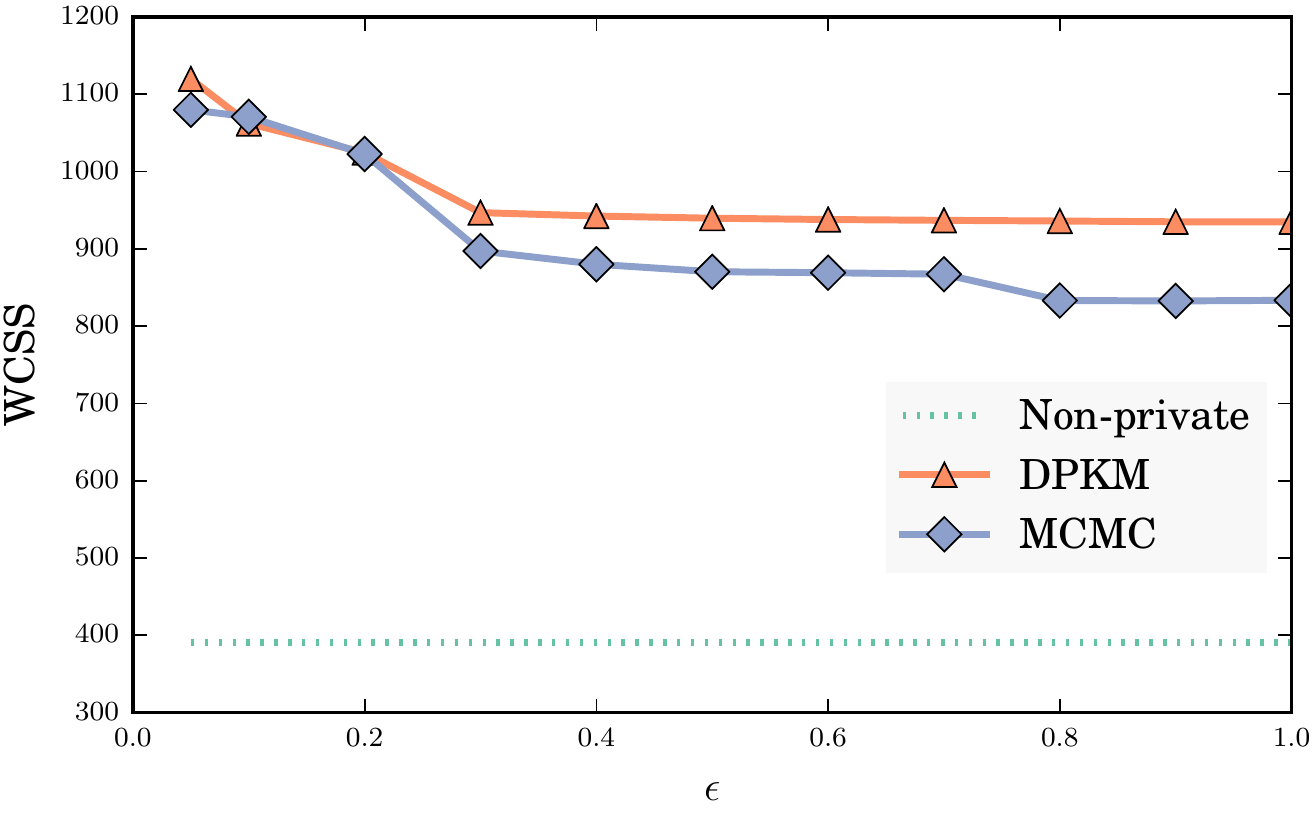}
    }
    \subfigure[Gowalla]{%
      \includegraphics[width=.3\textwidth]{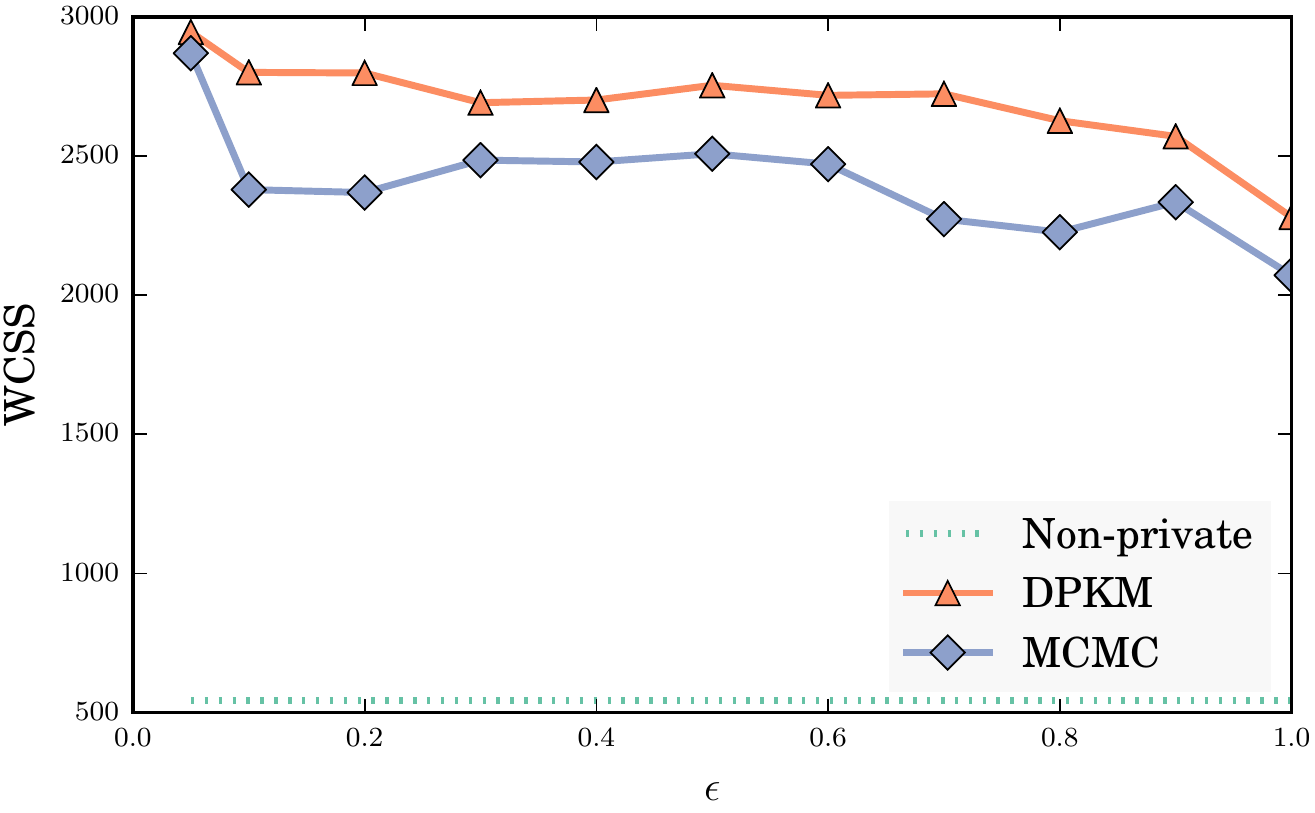}
    }
    \subfigure[Lifesci]{%
      \includegraphics[width=.3\textwidth]{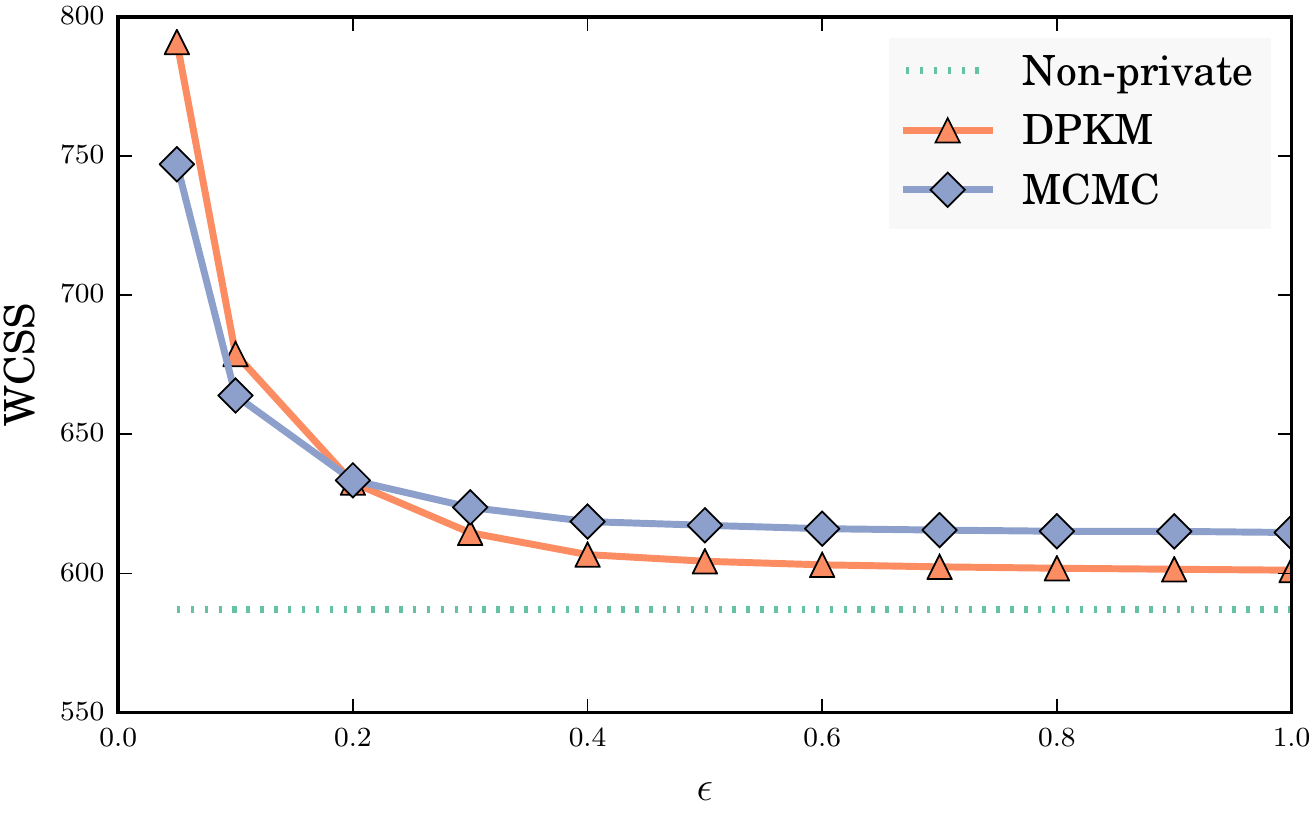}
    }
  \caption{WCSS by varying $\epsilon$}
  \label{fig:var_eps}
  \end{center}
\end{figure*}

Figure~\ref{fig:var_eps} shows the performance of our post-processing
algorithm for different values of $\epsilon$. The value of $\epsilon$
ranges from $0.05$ to $1.0$. The proposed algorithm improves the
accuracy of the final clusterings on all datasets, except on the Lifesci
dataset. On S1 and Tiger datasets, huge improvements in WCSS were
observed when $\epsilon=0.05$.


\section*{Acknowledgement}
This research was supported by NSF grant 1228669.

\bibliography{kmeans}
\bibliographystyle{icml2016}

\end{document}